\documentclass{sig-alternate-10pt}

\usepackage{subfigure}
\usepackage{times,url,comment,cite}
\usepackage{amsmath,amssymb,amsfonts,bbm}
\usepackage{graphicx,slashbox}
\usepackage{epsfig,epsf}
\usepackage{color}

\newtheorem{Definition}{Definition}

\title{A Lightweight Distributed Solution \\ to Content Replication
in Mobile Networks}
\numberofauthors{2} %  in this sample file, there are a *total*
\author{
\alignauthor
C.-A. La, P. Michiardi\\
       \affaddr{EURECOM}\\
       \affaddr{Sophia Antipolis, France}\\
       \email{firstname.lastname@eurecom.fr}
\alignauthor
C. Casetti, C.-F. Chiasserini, M. Fiore\\
       \affaddr{Politecnico di Torino}\\
       \affaddr{Torino, Italy}\\
       \email{firstname.lastname@polito.it}
}
\begin{document}
\maketitle

\begin{abstract}
Performance and reliability of content access in mobile networks is 
conditioned by the number and location of content replicas deployed at the network nodes. 
Facility location theory has been the traditional, \textit{centralized} approach to 
study content replication: computing the number and placement of replicas 
in a network can be cast as an uncapacitated facility location problem.

The endeavour of this work is to design a  \textit{distributed, lightweight}
solution to the above joint optimization  problem, while taking into account the
network dynamics.  In particular, we devise a mechanism that lets nodes share
the burden of storing and providing content, so as to achieve load balancing,
and decide whether to replicate or drop the information so as to adapt to a
dynamic content demand and time-varying topology.
%of establishing the number and location of content replicas to deploy in a mobile system. 
We evaluate our mechanism through simulation, by exploring a wide range 
of settings 
%including network and content consumption dynamics, 
and studying realistic content access mechanisms that go beyond the traditional assumption matching demand 
points to their closest content replica.
Results show that our mechanism, which uses \textit{local measurements only}, 
is: (i) extremely precise in approximating an optimal solution to content placement and replication; 
(ii) robust against network mobility; (iii) flexible in accommodating various content 
access patterns, including variation in time and space of the content demand.

\end{abstract}

\section{Introduction}
% \begin{itemize}
% \item Scope of this work
% \item The problam at a glance
%   \begin{itemize}
%   \item Why important (motivation)
%   \item Why difficult to solve
%   \end{itemize}
% \item Our approach and how it relates to the literature
% \item List of contributions (with forward references)
% \end{itemize}

%scope

%In recent years, the ability to access digital content published on the Internet
%has infringed the barrier of wireline connectivity as a result of the large
%penetration of mobile broadband technologies such as 3G and WiFi \cite{}. 
%Indeed, a steadily increasing number of mobile users access information
%wirelessly (e.g., podcasts, weather information, sports events) \cite{},
%creating the need for careful resource planning for mobile operators to avoid
%access congestion \cite{}.

Research activity in the networking field is pursuing the 
idea that networks should provide access to contents, rather than to
hosts. This idea is manifested in
content distribution networks based on either peer-to-peer networks,
% of hosts
% where contents are stored and retrieved based on distributed hash tables, 
or on an infrastructure of large storage nodes located close to
edge networks. In this paper, we explore this concept with respect to
wireless networks where nodes can exploit device-to-device communications. We highlight
that content is being stored in nodes that move and that the content
itself moves and is replicated in anticipation of being accessed. The purpose
is to re-assess content distribution with respect to wireless networks in
which content demand and topology are dynamically changing. 

Previous research has focused on techniques that enhance
performance and reliability of content access in wireless systems.
Content caching and replication have been shown  to be effective in
achieving these goals (see, e.g., \cite{Derhab09} for a survey on the topic). 
Every mobile device can potentially participate to
content caching or replication by storing data which can be made available to
other users through device-to-device communication over one of its wireless
interfaces, e.g., IEEE 802.11 or Bluetooth.

%devices with multiple communication interfaces are nowadays the norm rather than the exception \cite{}.

%problem: why hard and motivations

Due to the storage capacity constraints of cache servers, effective cache
eviction policies have traditionally been the subject of a large body of research
in computer science in general, and in the context of wireless networks
in particular \cite{cao06,das08,fiore09}. Decision problems concerning the
location of content replicas and, optionally, the number of replicas to deploy
in a network have also generated a large number of works: prominent examples of
such studies are available for wireline networks (e.g., content distribution
networks) \cite{Laoutaris07} and, to a lesser extent, for wireless networks \cite{Nuggehalli06}.

In this work we focus on \textit{content replication} in the context of mobile
wireless networks in which users create a cooperative environment. The very
nature of wireless content access and node mobility introduces several problems
to content replication.  {\em Optimal replica placement} is one of those:
selecting the location that is better suited to store content is difficult,
especially when the network is dynamic. However, optimal replica placement is
just a facet of content distribution: another prominent issue is {\em how many 
content replicas} should be made available to mobile nodes. Clearly, network and
content demand dynamics affect the solution of these two aspects. Furthermore,
decisions on the placement and number of replicas to be deployed in a network
are tightly related problems: intuitively, the latter introduces a feedback loop
to the former as every content replication triggers a new instance of the
placement problem.

% our approach (also vs. literature)
Traditionally, the above content replication problems have been studied through
the lenses of classic Facility Location Theory \cite{Mirchandani}: optimal placement can be
cast as the \textit{uncapacitated $k$-median} problem, whereas the joint
optimization of placement and number of replicas can be studied as an
\textit{uncapacitated facility location} problem. 
% The first problem assumes a
% fixed number $k$ of facilities to be placed so as to minimize the distance
% between demand points and their closest facility. Instead, in the second problem
% the number of facilities is not fixed, but jointly derived along with their
% location so as to minimize both the distances between demand points and their
% closest facilities, and the cost of opening (placing) a facility. 
Both these problems are
NP-hard for general network topologies; furthermore, the above formulations do
not tackle  the problem of {\em how users can access contents.}  

In our previous work \cite{casetti09}, we show preliminary results indicating
that a uniformly distributed replica placement  can be well  approximated using
distributed store-and-forward mechanisms, in which nodes store content only
temporarily. The endeavor of this work is to extend our previous study and
target the \textit{joint problem} (i) of establishing the number
of  replicas to deploy in a dynamic network, (ii) of finding their most
suitable location, and (iii) of letting users efficiently access stored content.
In particular, we address (i) and (ii) so as to achieve 
load balancing, that is, to let the network nodes evenly share
the burden of storing and providing content.

Instead of designing approximation algorithms of the optimal solution to
facility location problems which require global (or extended) knowledge of
the network \cite{arya01, Laoutaris07},  
we integrate our store-and-forward mechanism
with a distributed replication algorithm that bases its decisions on local
measurements only and aims at evenly distributing among nodes the demanding
task of being a replica provider. 
Also, we consider different content
query/reply mechanisms and evaluate their performance in conjunction with  the
schemes used for content replication and placement.

As a result, we show that both optimal placement and content replication can be
approximated through  a lightweight, distributed scheme which adapts to different
initial distributions of replicas and to variation in time and in space of
content demand,  while being robust against network dynamics.

% list of contributions with forward reference
%In summary, the main findings from our study are:
%\begin{itemize}
%\item 
%\end{itemize}
%{\color{red}PM: whould we put the paper organization?}

\section{Background and problem statement}
\label{sec:background}

As remarked above, the problem of content replication and caching has received
a lot of attention in the past due to its importance in enhancing performance,
availability and reliability of content access for Web-based applications. Here,
we inherit the problem of  replication typical of the wired-Internet and we
discuss why the dynamic nature of wireless networks  introduces new
challenges with respect to the wireline counterpart. Note that, although several
cache replacement policies have been proposed in the context of mobile ad-hoc
networks \cite{cao06,das08,fiore09}, in this paper we focus  on 
\textit{replication} and \textit{replica placement} problems, i.e., we view 
content replication as a process of its own, rather than
a by-product of a query/caching mechanism~\cite{Derhab09}. 

Let us now define the context of our work. We investigate a scenario involving
users equipped with devices  offering Internet broadband connectivity as well as
device-to-device communication capabilities (e.g., through IEEE 802.11).
Although we do not concern ourselves with  the provision of Internet access  in
ad hoc wireless networks, we remark that broadband connectivity is where  new
content is fetched from (and updated). 

In order to provide a basic description of the system,  we focus on content
being represented by a \textit{single} information object.  The mechanisms we
describe can then be extended to multiple objects. We assume the object to be
tagged with a validity time, and originally hosted on a server in the
Internet, which can only be accessed through the broadband access we hinted at. 
We then consider a \textit{cooperative network environment} composed of a set $V
= \{v(1), ..., v(N)\}$ of mobile nodes. A node $v(j)$ wishing to access the
content first tries to retrieve it from other devices; if its search fails,
the node downloads a fresh content replica from the Internet server and
temporarily stores it for a period of time $\tau_{v(j)}$,  termed
\textit{storage time}. For simplicity of presentation, in the following we
assume $\tau_{v(j)} = \tau,$  $\forall j\, \in V$. During the storage period,
$v(j)$ serves  the content to nodes issuing requests for it and, possibly,
downloads from the Internet server a fresh copy of the content if its validity
time has expired.  We assume that a node $v(i)$, which at a given time $t$  does
not store any copy of the content and which will later be referred to as
``content consumer'', issues queries at a rate $\lambda_{v(i)}(t)$. 

To achieve load balancing, at the end of the storage
time $v(j)$ has to decide whether (1) to hand the content over to another node,
(2) to drop the copy, or (3) to replicate the content and hand over both 
copies. We refer to the nodes hosting a content copy at a given time instant as
\textit{replica nodes},  and we denote their set by $\mathcal{C}(t)$.   
Only replica nodes are responsible for updating the content and for
injecting a new  version in the wireless network. 

%A couple of further considerations are required to clarify the network 
%dynamics in our work.
%\begin{itemize}
%\item \textit{Content transfer time versus storage time $\tau$.} 
%We assume the content size to be of the order 
%of few kilo-bytes. Due to the data transfer rates typical of 3G and IEEE 802.11 technologies, 
%we can safely assume transfer times to be of the order of few hundreds of milliseconds and, thus,  
%In our experiments, described in Section~\ref{sec:evaluation}, we set the storage time to be 
%of the order of hundres of seconds. Hence, we obtain the following inequality: 
%$\text{transfer time} << \tau$. Thus, the probability that the content expires
%during the transfer time can be neglected. 
%
%\item \textit{Content validity time $T_v$ versus inter-query time.} The time interval 
%between two subsequent 
%queries issued by a mobile node $v(i), \, \forall i \in V \setminus 
%\mathcal{C}(t)$, 
% must satisfy the following inequality: $1/{\lambda_{v(i)} (t)} \geq T_v$, $\forall t$, 
%for otherwise a user would obtain the same piece of information twice or more.
%\end{itemize}

Next, to highlight our contribution with respect to previous work,  we relate
our study to the formulation  of the replication and replica placement
problems typically used in the literature.  Let us fix the time instant and drop
the time dependency for ease of notation. Then, let $G=(V,E)$  represent the
network graph at the given time, defined by a node set $V$ and an edge set $E$. 
%Let $\lambda_{v(j)}$ be the (user)  content request rate originating from node
%$v(j)$. 
Let $\mathcal{C}$ denote the set of facility nodes,  i.e., nodes holding a
content replica. The specification of the placement of a given number of
replicas, $k$, amounts to solving  the uncapacitated $k$-median problem, which
is defined as follows.

\begin{Definition}
  \label{def:ukm}
  \textit{Uncapacitated $k$-median}. Given the node set $V$ with pair-wise 
  distance function $d$, service demand $\lambda_{v(j)}$, $\forall v(j) \in V$, select up to $k$ nodes 
  to act as facilities so as to minimize the joint cost $C(V,\lambda,k)$:
  \begin{equation}
    \label{eq:cost-ukm}
    C(V,\lambda,k) = \sum_{\forall v(j) \in V}{\lambda_{v(j)}d(v(j),m(v(j)))} \nonumber
  \end{equation} 
where $m(v(j)) \in \mathcal{C}$ is the facility that is \textit{closer} to $v(j)$.
\end{Definition}
The replica node set $\mathcal{C}$, instead, can be obtained by solving the 
following uncapacitated facility location problem at a given time instant. 
\begin{Definition}
  \label{def:ufl}
  \textit{Uncapacitated facility location}. Given the node set $V$ with 
  pair-wise distance function $d$, service demand $\lambda_{v(j)}$ and cost 
  for opening a facility at $v(j)$ $f(v(j)),$ $\forall v(j) \in V$, select a 
  set of nodes to act as facilities so as to minimize the joint cost $C(V,\lambda,f)$ 
  of acquiring the facilities and servicing the demand:
  \begin{equation}
    \label{eq:cost}
    C(V,\lambda,f) = \sum_{\forall v(j) \in \mathcal{C}}{f(v(j))} + \sum_{\forall v(j) \in V}{\lambda_{v(j)}d(v(j),m(v(j)))} \nonumber
  \end{equation} 
where $m(v(j)) \in \mathcal{C}$ is the facility that is \textit{closer} to $v(j)$.
\end{Definition}
For general graphs, both the above problems are NP-hard \cite{Kariv79} and a variety of approximation
algorithms have been developed, which however require global (or extended) knowledge of the 
network state \cite{arya01}.
%The problem of uncapacitated $k$-median assumes the size of the set $|\mathcal{C}| = k$ as an input to the problem, 
%hence (\ref{eq:cost}) reduces to the first term on the right hand side of the equation, i.e., to the 
%summation over the costs to open a facility.

Which new problems are introduced in the context of our work?  (i) Node mobility
introduces the problem of a dynamic graph $G$, requiring that  the facility
location problem be solved upon every network topology or demand rate 
change. (ii) Even under
static topology and constant demand, solving the facility location  problem does
not yield load balancing among nodes. (iii) The input to the facility location
problem is the content demand workload generated by users:  both replicas
location and the number of replicas to deploy in a network depend on content
consumption patterns.  While the  approach traditionally adopted  is to assume
content demand to be directed  to the closest facility, as stated in
Defs.~\ref{def:ukm}~and~\ref{def:ufl}, the wireless nature of our system
allows  content requests to propagate in the network, potentially reaching
multiple facilities (replica nodes).  

Our main contribution is therefore the design of a mechanism
for content placement and replication that achieves  load balancing as the
network topology and the query rate vary, while taking into account the
implications of query propagation towards replica nodes.

\section{Distributed mechanism for replication and placement problems}
\label{sec:system}

We now outline our content distribution and replication procedures.
% that are assumed to be used in our system. 
Firstly, several techniques for query distribution
and content access are detailed; next, we examine the challenging problem of
replica placement, i.e., of which nodes are to be selected as carriers of
content replicas to achieve load balancing;  finally, we discuss the behavior of replica nodes as a
function of  the system workload, in search of a cooperative,
distributed content replication strategy in presence of changing demand.

%We assume a \textit{cooperative environment} in which nodes
%the protocol execution and cache the content when necessary. 

%{\color{red} PM: missing an overview of what's next, and re-organize the
%assumptions we make.}

\subsection{Content access mechanisms}
\label{sec:content-access}

The workload experienced by a replica node is determined by the
mechanism used by nodes to access the content through device-to-device
communications. We identify two phases: a content query transmission, and 
a query reply transmission (by the replica node carrying the desired content).
We investigate several mechanisms for content
access focusing on the content query transmission phase, and we assume that
the identity of the nodes that have relayed the query is added to the 
query message itself. After
a replica node with the desired content is found, it will reply to 
the node issuing the query through a multihop transmission process
that backtracks the path from the replica node to the querying node, 
exploiting the identity of relay nodes included in the query message.
This backtracking, although possibly occurring
through multiple hops, makes no use of ad hoc routing protocols, as it is 
completely application-driven.

As far as the query transmission phase is concerned,  the following
three mechanisms are envisioned.

%\begin{itemize}
%   \item 
{\bf Scoped-flooding:} content requests are simply flooded
with a limited scope using application-layer broadcast. The ``scope'' can be
defined as the maximum number of hops through which a query propagates, i.e.,
neighboring nodes propagate a query until it has traversed a maximum number of
hops $H$, after which it is dropped.  Clearly, if the request is received by a
replica node, the content is served and the query is not propagated any further.

The main drawback of flooding is that multiple content replicas within reach of a
node will be ``hit'' by a request. 
Beside causing congestion when a large number of replica nodes reply
to the querying node, this also creates an artificially inflated
workload, which conflicts with the underlying assumptions in 
Defs.~\ref{def:ukm} and \ref{def:ufl}. 
In our experiments, we explore the benefits of a
\textit{selective reply} mechanism that replica nodes can use to mitigate
excessive workloads due to flooding. When selective reply is enabled, a replica
node replies to a query with a probability that is inversely proportional to the
hop-count of query messages.

%   \item 
{\bf Scanning:} instead of flooding in all directions, the
node issuing the query specifies an angular 
section within which the query is to be propagated by other nodes. 
In order to do so, it includes its own position (e.g., obtained through GPS),
and the angle boundaries. All nodes receiving the query rebroadcast it 
only if their position satisfies the angular requirements, until a replica 
node is found or the query has traversed a maximum number of hops $H$. 
Nodes that are not within the angular section specified in the query will 
discard the message. 
If no reply is received after a timeout, a new sector is scanned, and the 
scanning of all sectors is repeated till either a reply is received or
a maximum number of retries has been achieved. The number of  
sectors $S$, each of width $2 \pi / S$, is a parameter of the system.

The complexity of this mechanism is comparable to that of scoped-flooding,
however we will show
that it reduces the overhead experienced with flooding. On the downside,
scanning requires nodes to be able to estimate their position and
reduces the probability of solving a query with respect to flooding-based solutions.
Indeed, when a replica is within the sector currently scanned by
the requesting node but it is farther than one hop away, one or more relay nodes 
would be needed to reach the replica. However,
if at least one of the available relays are located outside the sector, the
replica is not reached and the content query remains unsolved. Thus, 
the narrower the sector, the more likely that the query is unsuccessful.
%   \item 

{\bf Perfect-discovery:} in this case, which is added for comparison 
purposes, nodes are assumed to be able to access a centralized content-location
service that returns the identity of the closest content replica in terms of
euclidean distance. We do not address the problem of how the centralized service
is updated, save by noting that it is certainly responsible for additional
overhead and complexity, and that it can be managed through a separate protocol using unicast
or multicast transmissions.
A query is propagated using application-driven broadcast, but
only the intended replica node (specified in the query) will serve the content.
Any other replica node will discard the request.

On the one hand, this content access mechanism is the most demanding because it
requires the presence of an auxiliary service to discover the closest replica.
On the other,  only one replica node carries the workload generated by the
closest users, which is the hypothesis to the optimization problems
stated in Sec.~\ref{sec:background}.

%\end{itemize}

%All access mechanisms above assume that a {\em query lag} is introduced  
%at each relay, to delay the propagation of a request in the hope that a 
%node in the neighborhood returns a response (thus making any further 
%query propagation useless). 

\vspace{10pt}

Finally, we improve the query/reply propagation process by adopting the PGB
technique \cite{naumov06} for selecting forwarding nodes and sequence numbers 
to detect and discard duplicate queries.

\subsection{Replica placement}
\label{sec:cache-placement}

Next, we overview the distributed lightweight algorithm that we use to
solve  the replica placement problem. 
% Recall that any mobile device can be
% selected for hosting a content replica, hence  the choice of a replica location
% is unconstrained. 
Recall that any mobile device can be selected to host a content replica 
for a limited amount of time, that we term \textit{storage time}, $\tau$.  
%hence  the choice of a replica location is unconstrained. 

Also, as the first step to our study, we focus on
the case of homogeneous user query rate, i.e., $\lambda_{v(j)}(t)=\lambda(t)$, 
$\forall v(j)$.    

As discussed in Sec.~\ref{sec:background}, at a fixed time instant, 
replica placement can be cast as
the uncapacitated $k$-median problem. Given a set of potential locations to
place a replica,  the problem is to position an a-priori known number $k$ of
replicas according to Def.~\ref{def:ukm}, i.e., so as to minimize 
the distance between replica node and requesting node. 
For a generic distribution of nodes over the network area, the solution 
of the $k$-median problem for different instances of the network graph
yields replica placements that are instances of a random variable 
uniformly distributed over the graph.
% so as to minimize the sum  of the distance between the replica and its
%closest nodes, which represent the demand points  of the problem. For instance,
%in case of a \textit{uniform distribution} of the nodes,  by solving the
%$k$-median problem for different instances of the network graph, we obtain that
%the optimal replicas placements are instances of a uniformly distributed random
%variable.  
This is quite an intuitive result, confirmation of which we found by 
applying the approximation algorithm in \cite{arya01} to the solution of the
$k$-median problem in presence of various network deployments.  
%As a consequence, when nodes are uniformly distributed in space, the ideal placement of caches 
%is also uniform.
%Similarly, in the case of a generic distribution of the nodes over the network
%area, for different instances of the deployment, the optimal replicas
%placements  result to be instances of a random variable that is uniformly distributed
%over the nodes graph, i.e., the optimal replica placement reflects the node
%deployment. 
%We refer to the latter as {\em nodal uniformity}.

%In our case, however, 

As pointed out earlier, though, the solution in \cite{arya01} cannot be applied  to
our case since it is centralized and requires global knowledge of the network. 
%Our approach to approximate the optimal replica placement targets, instead,
%a uniform distribution of the replicas  over the network nodes, i.e., a
%\textit{nodal uniformity}.  
We therefore devise a lightweight distributed mechanism that
well approximates  a uniform distribution of the replicas  over the network nodes, i.e., a
\textit{nodal uniformity},  
%target nodal uniformity of replica locations when the network graph is dynamic, 
and that allows users to take turns in playing the role of
replica nodes so as to achieve load balancing. 

% Recall that any mobile device can
% be selected to host a content replica for a limited amount  of time, that we
% term \textit{storage time}, $\tau$. 
According to our mechanism, named  \textit{Random-Walk
Diffusion (RWD)}, at the end of its storage time, a replica node 
selects  with
equal probability one of its neighbors to store the content for the following
storage period.  Thus, content replicas roam the network by moving from one node
to another, randomly,  at each time step $\tau$.  

%Note that the rational of
%resorting to a random walk process relies on the following result \cite{CRW07}:
%if the network topology can be represented as an undirected,  connected,
%non-bipartite graph, then the distribution of  nodes moving according to the
%random walk model converges to a unique stationary distribution regardless of
%the initial distribution, and  this stationary distribution is uniform  over
%nodes in the case of regular graphs\footnote{A graph is regular if each of its
%vertices has the same  number of neighbors.}. 

% In \cite{casetti09}, we also
% explored a more elaborate store-and-forward  algorithm, called Random-direction
% Diffusion, and compare its performance against RWD in  a variety of scenarios.

To understand the extent to which replica placement achieved by our simple technique resembles 
the target nodal distributions, in Sec.~\ref{sec:evaluation}  we employ the well-known  $\chi^2$
goodness-of-fit test on the inter-distance between content replicas.  Whenever the computation complexity
allows us, we compare the temporal
evolution of the inter-distance distribution of replicas obtained by our scheme against the optimal
replica placement computed by solving the $k$-median problem. Otherwise, we consider as term of comparison
the empirical distribution of the distance between two nodes measured in simulation.
Note that using inter-distances instead of actual coordinates
allows us to handle a much larger  number of samples 
(e.g., $|V|\cdot(|V|-1)$ instead of just $|V|$ samples)
thus making the  computation of the $\chi^2$ index more accurate.

It is clear that the quality of approximation of the target replica distributions achieved by our
store-and-forward mechanism depends on the node density: 
the higher the density, the better our 
approximation.  
%Furthermore, as observed in the preliminary analysis of our mechanism
%\cite{casetti09},  the approximation quality is positively correlated to the total number of
%replicas to place. 
% Next, we describe how we address the problem of determining the number of replicas
% in the network and how we design a mechanism that is able to adapt to both temporal
% and spatial variations in the content demand.

\subsection{Content replication}
\label{sec:content-replication}

%The underlying assumption to the replica placement problems is an a-priori knowledge of the number
%of replicas ($k$) that needs to be placed in the network.  
We now focus on the more general
problem of the uncapacitated facility location,  defined in Sec.~\ref{sec:background}, where the
optimal number of replicas (facilities) to be placed in the network  is to be determined along with their location. 
In particular, we want to answer the following questions.

\begin{enumerate} 
\item Given a set of demand points that exhibit a
homogeneous query rate $\lambda$, what is the optimal number 
of content replicas that should be
deployed in the network to achieve load balancing?  
\item Is it possible to design a lightweight
distributed algorithm that approximates this optimal number of replicas  in presence of a dynamic
demand and time-varying topology?  
\end{enumerate} 

We address these questions by suggesting simple modifications
to the RWD mechanism described in Sec.~\ref{sec:cache-placement}. 

Again, we fix the time instant and, for simplicity, we drop the time dependency from our notation. 
Let the network be described by the graph $G=(V,E)$, with $|V|=N$ nodes deployed 
on an area $\mathcal{A}$. Also, recall that $\mathcal{C}$ and $V \setminus \mathcal{C}$ 
represent the sets of content replicas and of nodes issuing requests, respectively. 

Given $G$ and the query rate $\lambda$, the uncapacitated facility location
problem amounts  to the joint optimization of the number of replicas and their
locations in the network.  The RWD mechanism achieves a good approximation of
the optimal placement in mobile networks,  but ignores the cost to deploy a
content replica. Now, with reference to  Def.~\ref{def:ufl}, we define the cost
function to deploy content replicas in the network 
$f(v(j))$, $\forall v(j) \in \mathcal{C}$, as follows:
\begin{equation}
  \label{eq:open-cost}
  f(v(j)) = | s_{v(j)} - s_{R} |
\end{equation}
where $s_{v(j)}$ is the workload expressed as number of queries served by 
replica node $v(j)$ during its storage time,
and $s_{R}$ is a reference value for the workload that node $v(j)$ is willing 
to support. We assume the case where all replica nodes are willing 
to serve the same amount of queries, although our study can be easily extended to the case of different
values of $s_R$. 
Eq.~(\ref{eq:open-cost}) indicates that the cost for replica node $v(j)$ grows 
with the gap between its workload and the reference value $s_{R}$. 
By using the cost function in (\ref{eq:open-cost}) in the facility location
problem in Def.~\ref{def:ufl}, we can determine the location and  
number of replicas so that load balancing is achieved under the idealistic 
assumption that each query reaches one replica only. 
Our replication mechanism only involves replica nodes, 
which are responsible to decide whether to replicate, hand over or drop content based on local measurements 
of their workload. 
%
%As before, we focus on a homogeneous case 
%in which all nodes expect the same reference workload $s_R$.
%{\color{red}PM: I've noticed that in this work we say nothing about the impact of $s_{R}$ 
%on replication, how to determine this values, nor we look at an heterogeneous case.}
During storage time $\tau$, the generic  replica node $v(j)$ \textit{measures} 
the number of queries that it serves, i.e., $\hat{s}_{v(j)}$. 
When the storage time expires, 
%if the content copy is still valid,
the replica node compares $\hat{s}_{v(j)}$ to $s_{R}$. Decisions are taken as follows:
\[
\text{if  }
\hat{s}_{v(j)} - s_{R} 
\left\{
\begin{array}{ll}
> \epsilon \quad & \text{replicate } \\
< - \epsilon \quad &\text{drop }\\
 \text{else} \quad &  \text{hand over}
\end{array}
\right.
\]
where $\epsilon$ is a tolerance value to avoid replication/drop decisions 
in case of small changes in the node workload.

The rationale of our mechanism is the following. If $\hat{s}_{v(j)} > s_{R}$,
replica node  $v(j)$ presumes the current number of content replicas in the area
to be insufficient to  guarantee the expected workload $s_{R}$, hence the node
replicates the content and hands   the copies over two of its neighbors (one
each), following the RWD placement mechanism
(Sec.~\ref{sec:cache-placement}). The two selected neighbors will act as
replica nodes for the subsequent storage time.  Instead, if $\hat{s}_{v(j)} <
s_{R}$, replica node $v(j)$ thinks that the current number of  replicas in the
area is exceeding the total demand, and just drops the content copy. Finally, if
the experienced workload is (about) the same as the reference value, $v(j)$ 
selects one of its neighbors to hand over the current copy. 

We stress
that replication and placement are tightly related.  For example, if content
demand varies in time or in space (e.g., only a fraction of all nodes located
in  a sub-zone of the network area issue queries), both the number of replicas
and their location must change.  Thanks to the fact that replica nodes take
decisions based on the measured workload, our solution can dynamically
adapt to a time- or space-varying query rate, as will be shown by our simulation
results. On the contrary, when the content demand is constant and homogeneous,
our handover mechanism ensures load balancing among the network nodes.

In the following, we set up a simulation environment to evaluate the behavior 
of our mechanism when the wireless network is both static and dynamic. 
We also characterize the time the system takes to reach an optimal number of
content replicas and 
we investigate the impact of the content access scheme on the performance 
of our solution.
%ci aspettiamo quando il perfect, sia vicino all'ottimo,
%scannig e ancora peggio flloding -based, no. di repliche aumenti.
\section{Simulation-based evaluation}
\label{sec:evaluation}

We implemented our replica placement and content replication mechanism in the
$ns$-2 simulator. For each experiment described in the following, we execute 10
simulation runs and report averaged results. Our statistics are collected after
an initial warm-up period of 500 s.

In our simulations, which lasted for almost 3 hours of simulated time (10000 s),
we assume nodes to be equipped with a standard 802.11 interface, with an 11 Mbps
fixed data transmission rate and a radio transmission range of 20 m.  We
consider a single content, whose size is of the order of 1 KB. In our evaluation
we do not simulate cellular access. We point out that all standard MAC-layer
operations are simulated, which implies that both queries and replies may be
lost due to  typical problems encountered in 802.11-based ad hoc networks
(e.g., collisions or hidden terminals).  This
explains why, in the following, even nominally ``ideal'' access techniques may
not yield the expected good performance.

We focus our attention on wireless networks with high node density: we place
$N=320$ nodes uniformly at random on a square area $\mathcal{A}$ of 
$200 \times 200$~m$^2$, with a resulting average node degree of 9--10 neighbors. We simulate node
mobility using the \textit{stationary} random waypoint model ~\cite{leboudec05}
where the average node speed is set to 3 m/s and the pause time is set to 100
s. These settings are representative, for example, of people 
using their mobile devices as they walk.

Unless otherwise stated, the parameters that define the mechanisms described in
this work are set as follows. For the content access mechanisms, we set the
scope of flooding and scanning to $H=5$ hops: e.g., a node can cover half of the
network diameter with scoped-flooding.  In the case of scoped-flooding or
perfect-discovery, if a query fails (i.e., no answer is received after 2 s), a
new request is issued, up to a total of 5 times. If the scanning mechanism is
used, a complete scan of $2\pi$ is divided into $S=5$  angular sectors, each of
which is visited  for a maximum of 0.5 s, at most 5 times\footnote{We use
the parameters that give the best results in terms of content access
performance.}.

Finally, the tolerance value $\epsilon$ used in the replication/drop algorithm is equal
to 2, unless otherwise stated; for all nodes, the storage time $\tau$ is set to 100 s, the user
request rate is $\lambda=0.01$ req/s, and the reference workload for a
replica node is equal to $s_{R}=10$.

% In this work we present the most significant results for the RWD replica placement mechanism: a detailed analysis of can be found in \cite{casetti09}, where we explore a large fraction of the parameter space that characterize the behavior of our randomized schemes.

\subsection{Results}
\label{sec:results}

We present the main results of our work organized in a series of questions. 
We focus on the mobile scenario, but
present results for a static network when the comparison is relevant.
Tab.~\ref{tab:content_access} summarizes the notations used in our figures to
refer to content access mechanisms.

\begin{table}[htbp]
 \centering
\caption{\label{tab:content_access} Notation for different content 
  access mechanisms. The post-fix ``S'' and ``M'' 
  indicate a static and a mobile network, respectively.}
  \begin{tabular}{ll}
    \hline 
    {\textit{Content access mechanism}} & {\textit{Notation}} \\
    \hline 
    {Perfect-discovery} & {PS, PM} \\
    {Scanning} & {SS, SM} \\
    {Scoped-flooding} & {FS, FM}\\
    {Scoped-flooding with selective reply} & {FS*,FM*}\\
    \hline
  \end{tabular}
\end{table}

\subsubsection*{How well does our replica placement approximate the target
distribution?}

Here we assume a known number of content replicas to be deployed
($|\mathcal{C}|$=30), i.e., we consider the $k$-median problem 
discussed in Sec.~\ref{sec:background}. We
measure the accuracy of our distributed replica placement mechanism using the
$\chi^2$ goodness-of-fit test on the inter-distance between replicas, as
explained in Sec.~\ref{sec:cache-placement}. In case of a mobile network, we
compute the distribution of replica nodes as follows: every $\tau$ seconds we
take a snapshot of the network in its current state, we compute a reference
distribution (e.g., nodal uniformity) of content replicas and use the $\chi^2$
test against the distribution achieved by our mechanism. We remark that
lower values of the $\chi^2$ index indicate a better approximation, and 
that usually a value of $\chi^2$ below 3.84 is considered a good fit \cite{nist}.

\begin{figure}[htbp]
  \centering
  \includegraphics[scale=0.3]{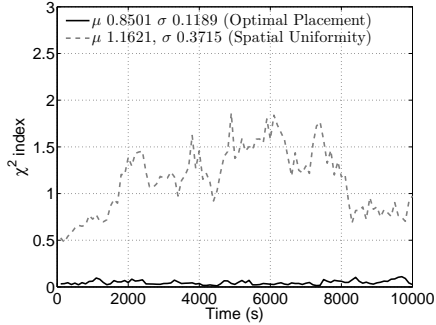}
   \caption{Temporal evolution of the $\chi^2$ index 
   in a static scenario ($|\mathcal{C}|$=30 and $\tau$=100 s).}
   \label{fig:static_chi}
\vspace{-3mm}
\end{figure}

We first focus on a static network in which nodes are uniformly distributed on
$\mathcal{A}$. 
Fig.~\ref{fig:static_chi}  shows that our scheme  does an excellent good job of
approximating the optimal replica placement\footnote{In fact, we solve the
$k$-median problem through the centralized local-search algorithm described in
\cite{arya01} and obtain a tight approximation of the optimal solution.}.  
(We omit the comparison to nodal uniformity since our results show
that the values of  the $\chi^2$ index are practically identical.) We
therefore asked ourselves if  a similar match could be found if the replica
placement were uniformily  distributed in space. As can be seen in the figure, a
higher value of  $\chi^2$ indicates a poorer match with our placement scheme,
mainly due to the  lack of nodes (hence of replicas) where the spatially uniform
distribution  would have theoretically placed them. 

\begin{figure}[htbp]
  \centering
  \includegraphics[scale=0.3]{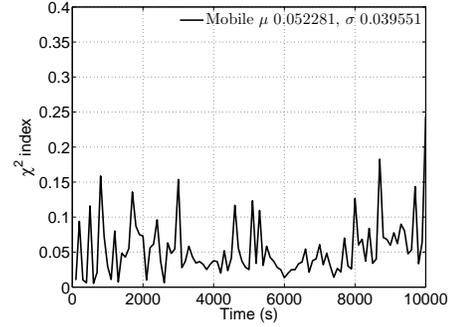}
   \caption{Temporal evolution of the $\chi^2$ index in a mobile scenario ($|\mathcal{C}|$=30 and $\tau$=100 s).}
   \label{fig:mobile_chi}
\vspace{-3mm}
\end{figure}

Fig.~\ref{fig:mobile_chi} depicts the $\chi^2$ test against \textit{nodal
uniformity} for the mobile scenario. Note that, due to the \textit{stationary}
random way-point mobility model used in our simulations, the node  distribution
is not uniform on the network area. The temporal evolution of the $\chi^2$ index
suggests that our replica placement mechanism is able to approximate very well
nodal uniformity, despite network dynamics. We did not consider the optimal
placement in this case, due to the cumbersome computational load.

\begin{figure}[htbp]
  \centering
  \includegraphics[scale=0.3]{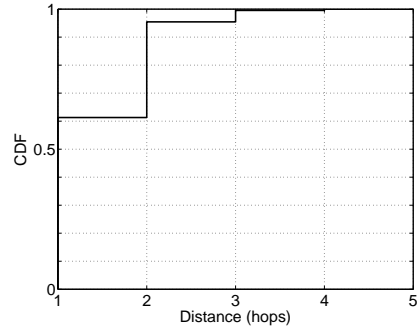}
   \caption{CDF of the hop-distance to the closest content replica in a mobile scenario ($|\mathcal{C}|$=30 and $\tau$=100 s).}
   \label{fig:mobile_distance}
\vspace{-3mm}
\end{figure}

Finally, Fig.~\ref{fig:mobile_distance} reports the cumulative distribution function
(CDF) of the distance (in number of hops) between a content consumer and
the closest replica. We observe that more than 50\% of nodes can reach the closest replica
within 1 hop, and more than 90\% of nodes are within 2 hops from the closest
replica, i.e., for $|\mathcal{C}|$=30, consumers can access content 
within very few hops.

\vspace{10pt}

\textit{
\textbf{Summary}: we overviewed our replica placement 
mechanism as a necessary introductory step to the replication scheme. 
We showed that the RWD mechanism can approximate very accurately the 
optimal solution to the $k$-median problem (Fig. \ref{fig:static_chi}), and it clearly 
approximates nodal uniformity even when 
the network is dynamic (Fig. \ref{fig:mobile_chi}). For the placement problem alone, in \cite{casetti09} 
we also explored the implications of clustered networks, different 
mobility models, and  the parameter $\tau$. 
}

\subsubsection*{What is the performance of content access mechanisms?}
We evaluate the performance of the four content access mechanisms 
listed in Tab.~\ref{tab:content_access}, in terms of
the following metrics:

\begin{itemize}

\item {\it solving ratio}, i.e., the ratio of satisfied requests to the
total number of queries generated in the network.  The target value is $1$,
corresponding to 100\% of solved queries;

\item {\it reply redundancy}, i.e., the number of replies to the same request,
received from different replica nodes. The target value is $1$, corresponding to
one reply to each query;

\item {\it latency}, i.e., the delay experienced by nodes to access
information.

\end{itemize}

\begin{figure*}[htbp]
  \begin{tabular}{ccc}

    \begin{minipage}[t]{0.3\textwidth}
      \begin{center}    
        \subfigure[Solving ratio]{
          \label{fig:sat_bp}
          \includegraphics[scale=0.3]{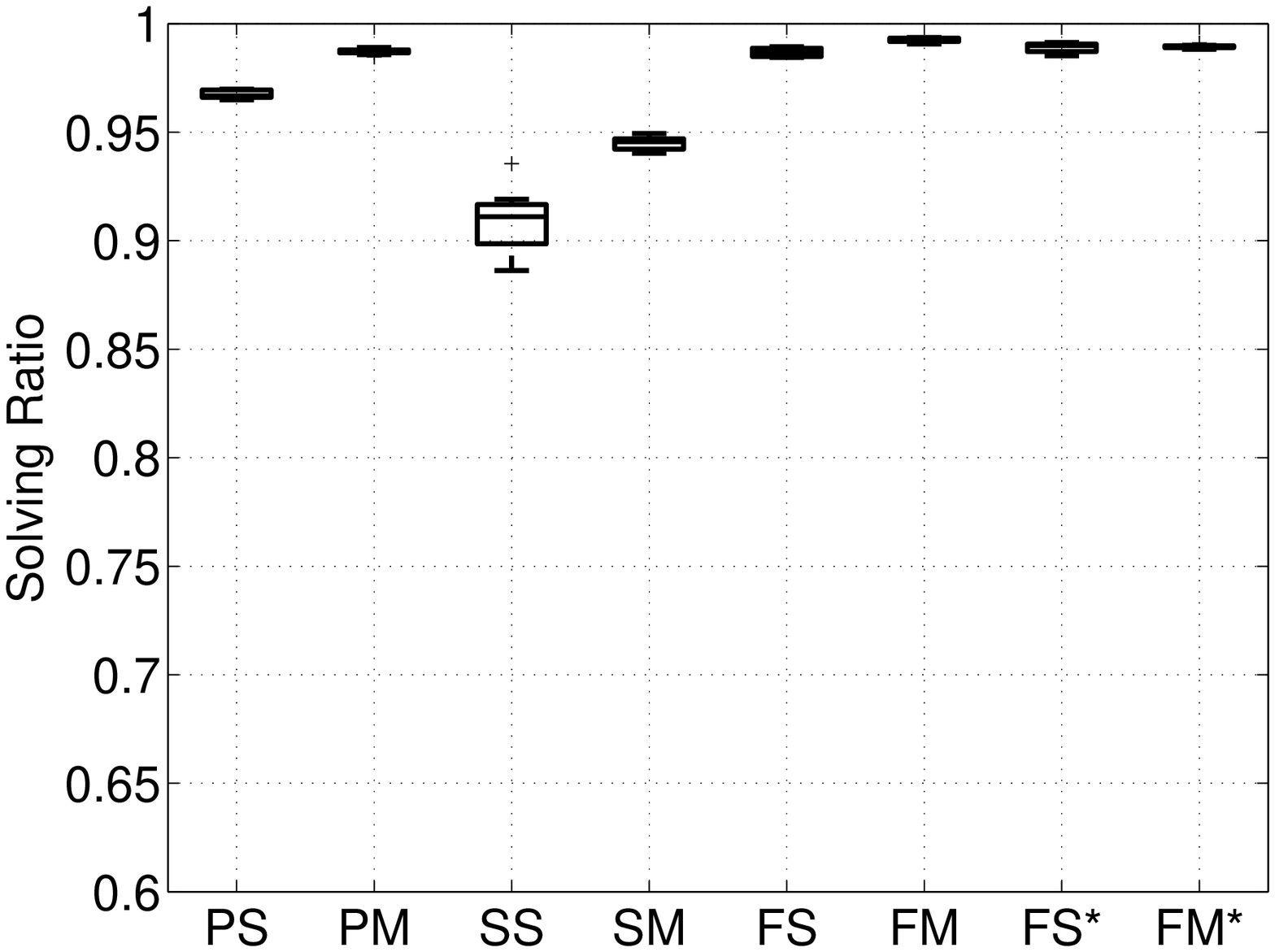}}
      \end{center}
    \end{minipage}

    &

    \begin{minipage}[t]{0.3\textwidth}
      \begin{center}  
        \subfigure[Reply redundancy]{
          \label{fig:dup_bp}
          \includegraphics[scale=0.3]{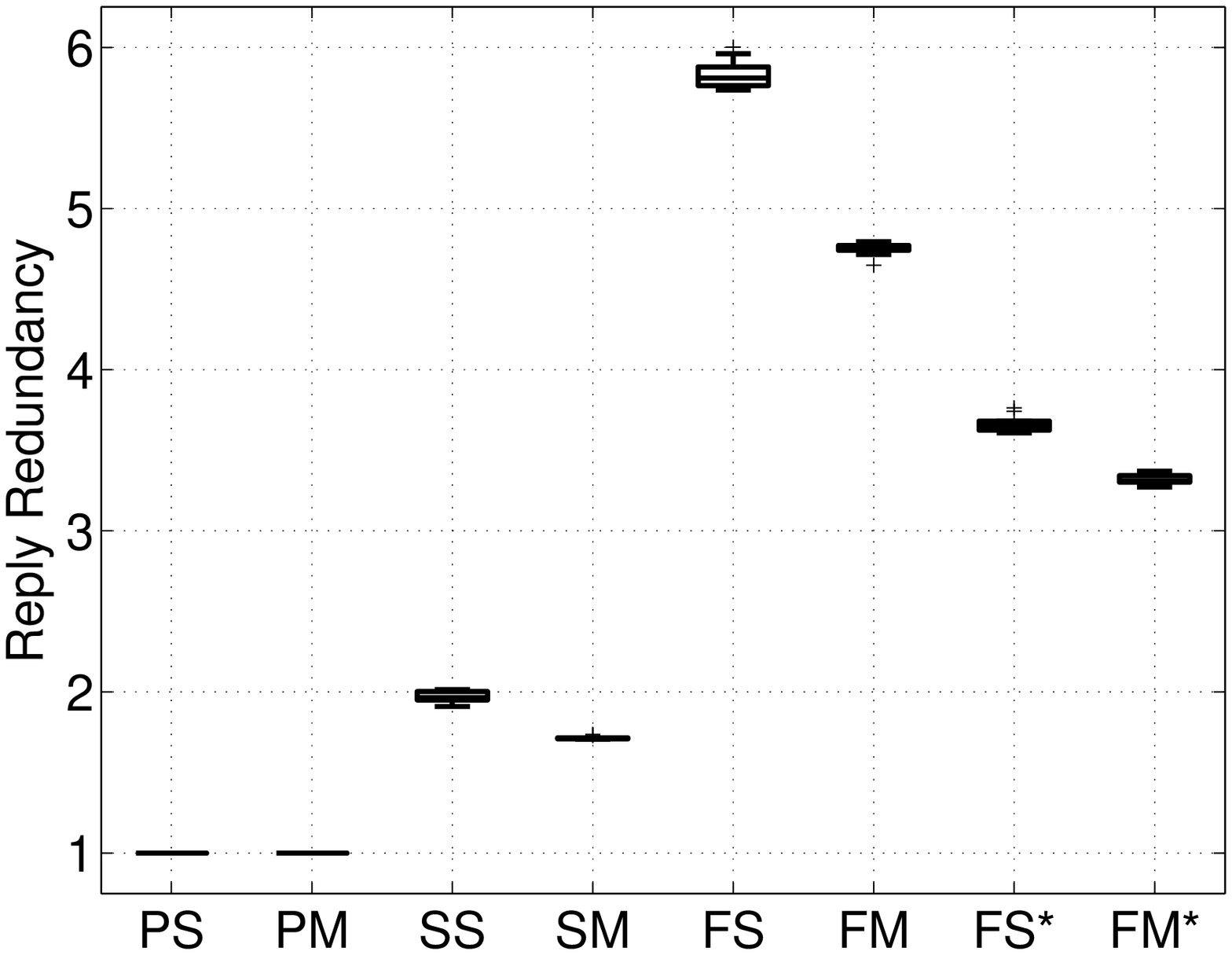}}
      \end{center}
    \end{minipage}

   &

    \begin{minipage}[t]{0.3\textwidth}
      \begin{center}    
        \subfigure[Latency]{
          \label{fig:delay_bp}
          \includegraphics[scale=0.3]{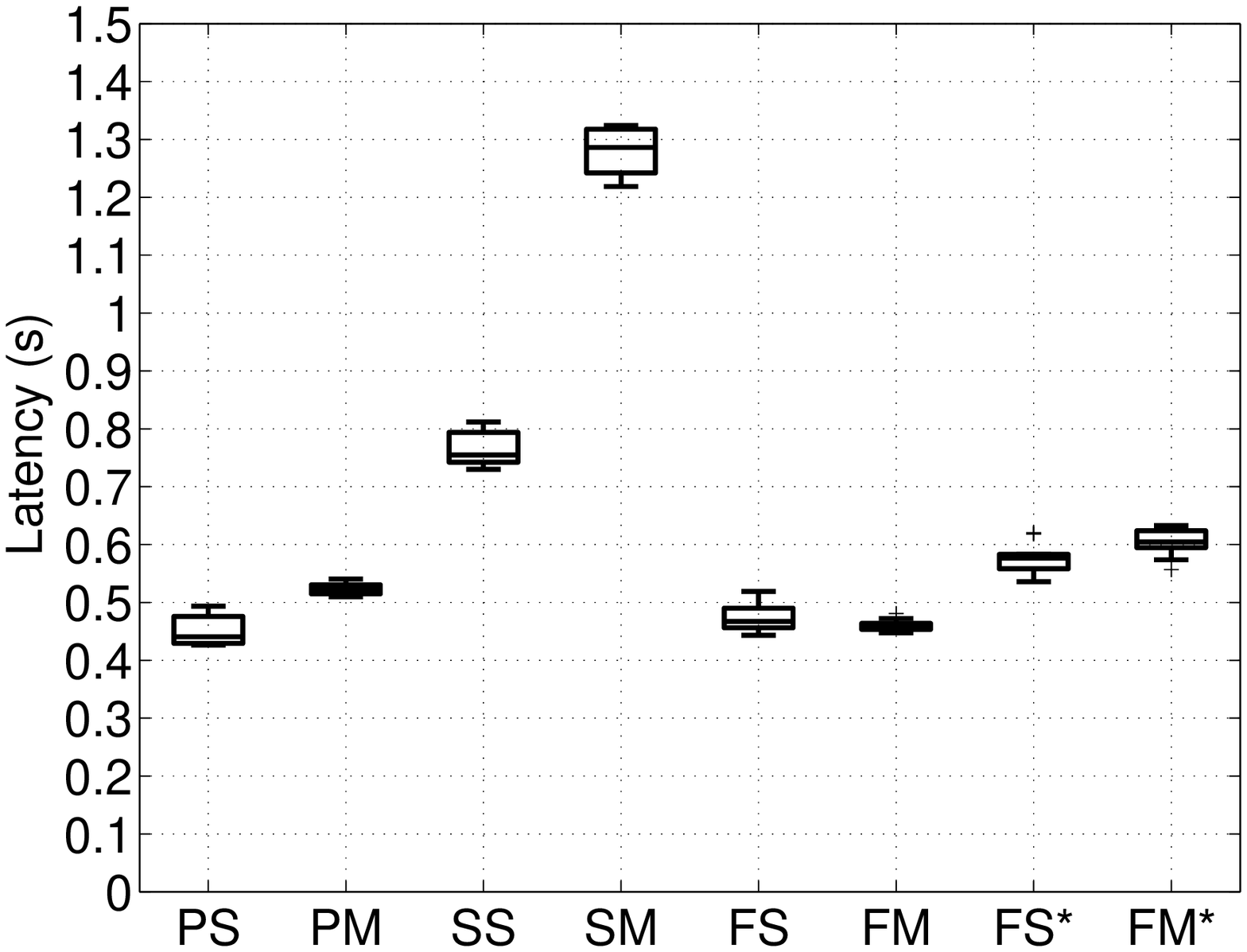}}
      \end{center}
    \end{minipage}
    
  \end{tabular}
  \caption{Performance of content access mechanisms, in a static and mobile scenario ($|\mathcal{C}|$=30 and $\tau$=100 s).}
  \label{fig:performance_bis}
\vspace{-3mm}
\end{figure*}

Fig.~\ref{fig:performance_bis} shows the following quantiles of the
access performance metrics for $|\mathcal{C}|$=30: the 25\% (resp. 75\%) as
the lower (resp. higher) boundary in the error box, the 50\% as the 
line within the error box. The brackets above and below the error box 
delimit the support of the CDF for that metric.
For all access mechanisms, the median solving ratio
(Fig.~\ref{fig:performance_bis}.a) is higher than 0.9, which indicates that only
a small fraction of queries cannot reach a content replica. We observe that node
mobility helps improving the solving ratio. The scheme that 
exhibits a slightly worse performance appears to be the scanning scheme, 
which is seldom unabke to reach a replica through relay nodes (see Sec. \ref{sec:system}).
%smaller timeout before declaring the query as unsolved.
%However, it should be noted that such timeout choice allows the 
%scanning scheme to trade off a lower solving ratio with a smaller latency. 
 
Fig.~\ref{fig:performance_bis}.b depicts the extent to which flooding-based
mechanism can artificially inflate the workload of replica nodes: in our
experiments, a single query can hit almost 6 replicas in the worst 
case\footnote{We also run
experiments with lower values of the flooding scope $H$: redundancy, which is
proportional to $H$, remains higher in flooding compared to other schemes.}. High
redundancy has a direct consequence on the behavior of the replication
mechanism, as we discuss in detail later. We observe that the selective reply
mechanism can halve the level of redundancy typical of flooding, and that node
mobility helps in reducing redundancy in all schemes. It is important
to notice that the scanning mechanism achieves a low reply redundancy, without
requiring the presence of an auxiliary mechanism to help consumer nodes
target the closest replica.

Latency for each content access scheme is shown in
Fig~\ref{fig:performance_bis}.c: 
scanning is clearly the outlier in this figure,
and we observe that node mobility might introduce additional delays.

\begin{figure*}[htbp]
  \begin{tabular}{ccc}

    \begin{minipage}[t]{0.3\textwidth}
      \begin{center}    
        \subfigure[Solving ratio]{
          \label{fig:sat-scan-time}
          \includegraphics[scale=0.3]{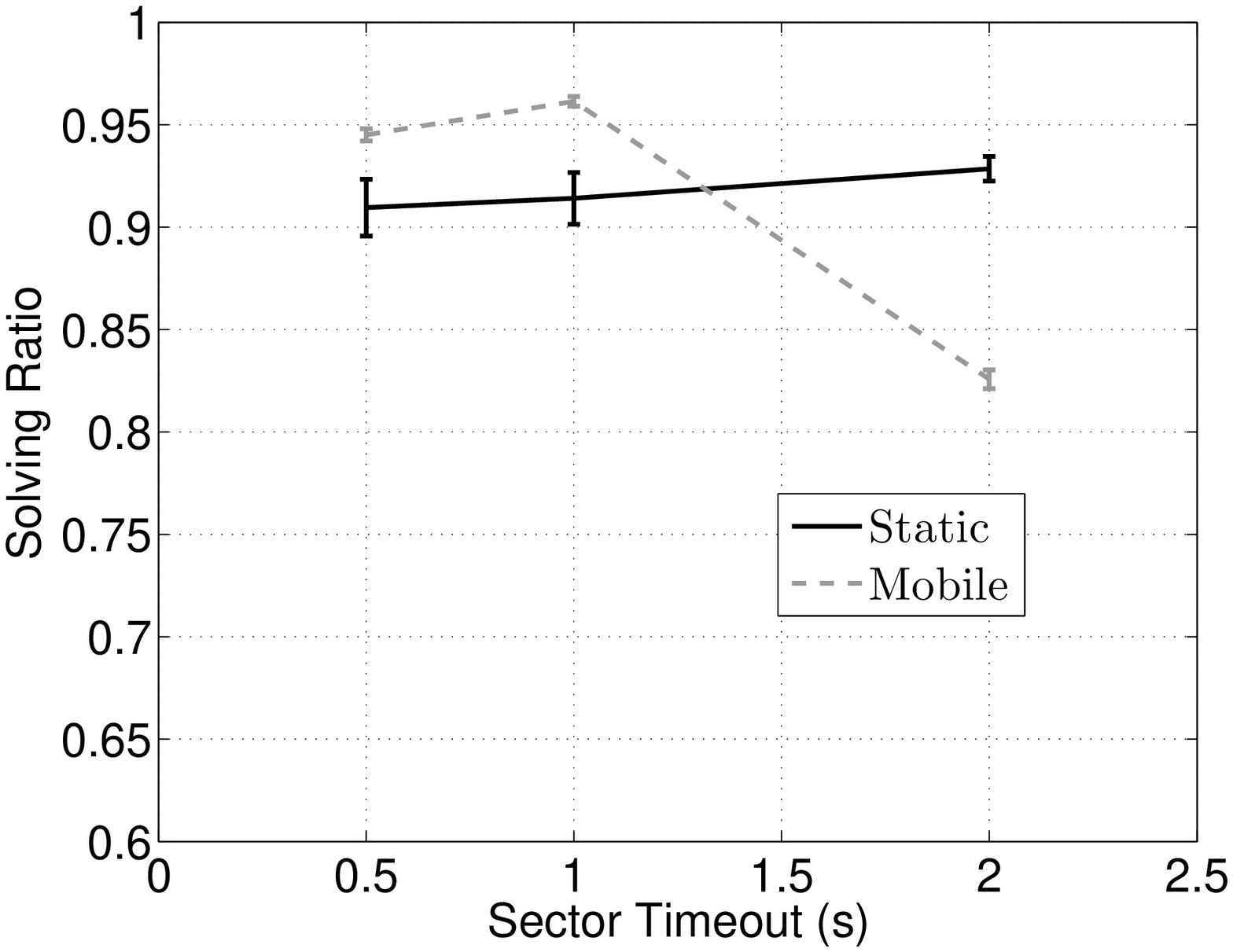}}
      \end{center}
    \end{minipage}

    &

    \begin{minipage}[t]{0.3\textwidth}
      \begin{center}  
        \subfigure[Reply redundancy]{
          \label{fig:dup-scan-time}
          \includegraphics[scale=0.3]{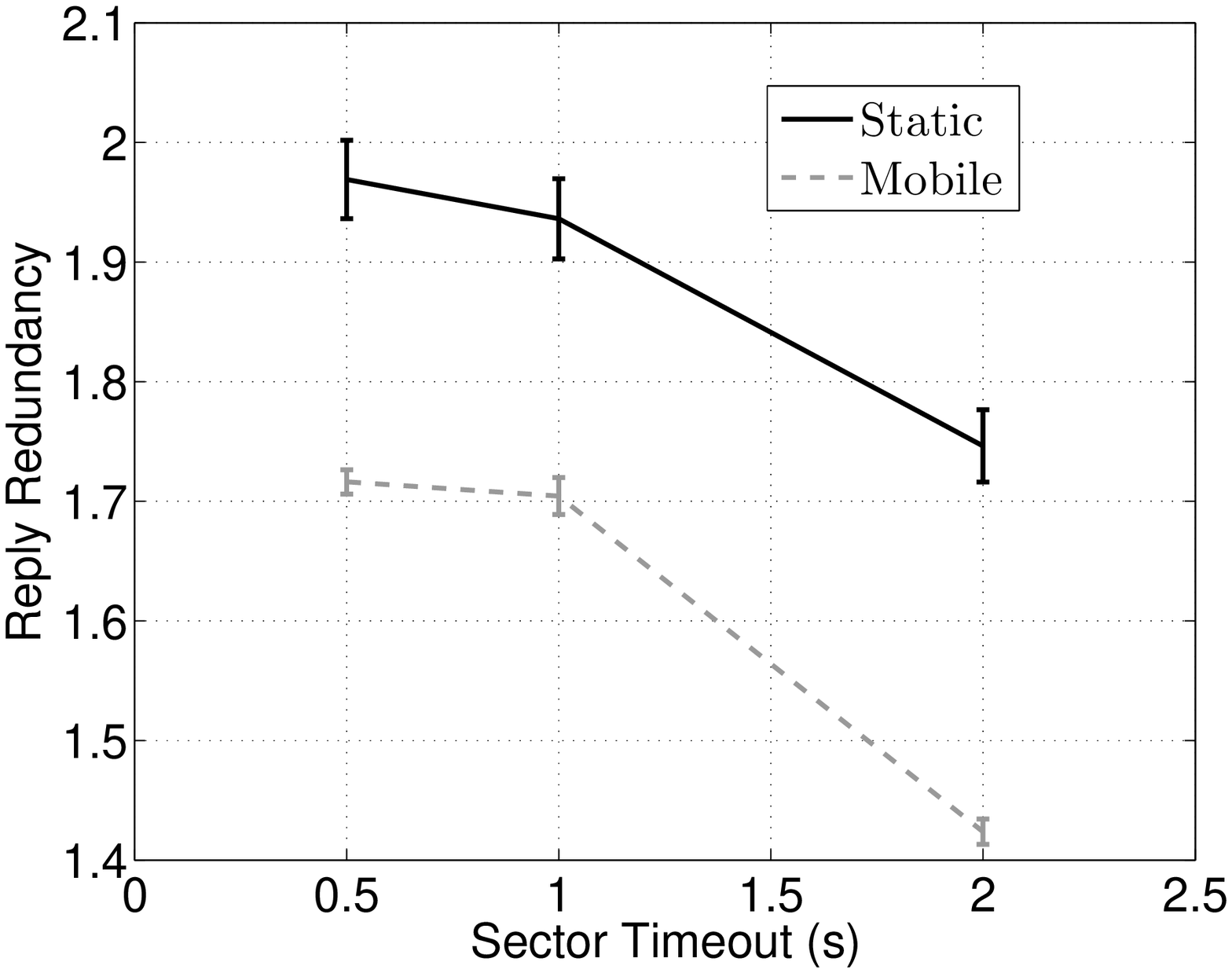}}
      \end{center}
    \end{minipage}

   &

    \begin{minipage}[t]{0.3\textwidth}
      \begin{center}    
        \subfigure[Latency]{
          \label{fig:delay-scan-time}
          \includegraphics[scale=0.3]{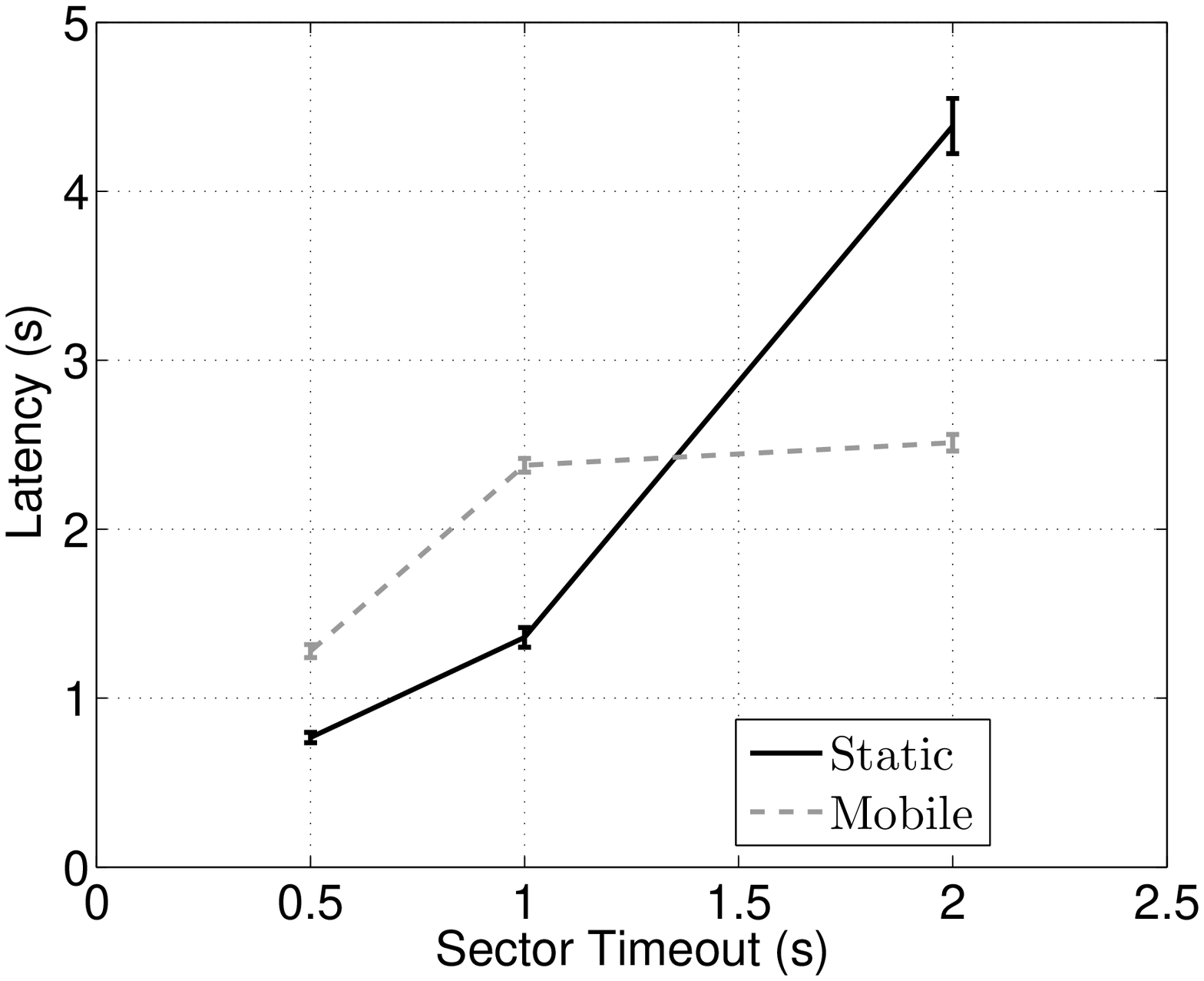}}
      \end{center}
    \end{minipage}
    
  \end{tabular}
  \caption{Performance of the scanning mechanism as a function of the sector timeout (scanning angle $2 \pi / 5$, $|\mathcal{C}|$=30 and $\tau$=100 s).}
  \label{fig:performance-scan-time}
\vspace{-3mm}
\end{figure*}

We now provide more details on the performance of the scanning mechanism.
Fig.~\ref{fig:performance-scan-time} shows the impact of the time spent waiting for a reply 
on each sector composing the scanning horizon; we term this time \textit{sector timeout}. 
The solving ratio is marginally affected by this parameter for
a static network, but it decreases in the mobile case. Indeed,
delaying the search in the next sector by a longer time has the mobile
node skip larger portions of the area: two consecutively scanned sectors turn out to be 
non-adjacent due to the change in the position of the node issuing the query.
The redundancy decreases
with longer sector timeouts: indeed, the longer the timeout the higher the probability 
that a node scans another sector, i.e., it issues another query, only when no replica is 
available in the current sector.  Instead,
the latency deteriorates with a longer sector timeout because it will take more
time to hit the sector where the replica is located.  
Mobility seems to have a positive effect on the delay, even with
longer sector timeouts, since most solved queries are due to close-by 
replica nodes (farther nodes may reply after the querying node has moved away).

\begin{figure*}[htbp]
  \begin{tabular}{ccc}

    \begin{minipage}[t]{0.3\textwidth}
      \begin{center}    
        \subfigure[Solving ratio]{
          \label{fig:sat-scan-angle}
          \includegraphics[scale=0.3]{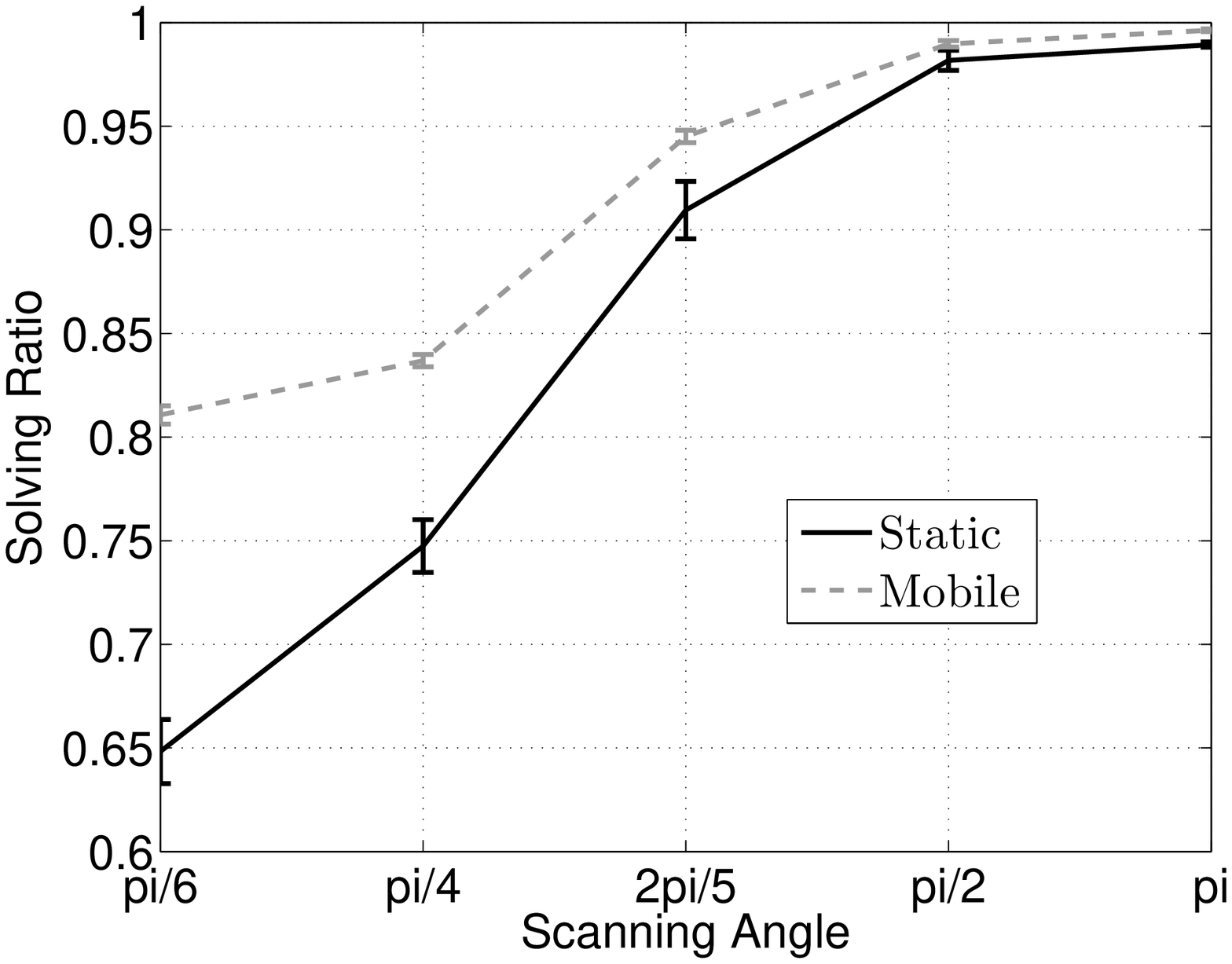}}
      \end{center}
    \end{minipage}

    &

    \begin{minipage}[t]{0.3\textwidth}
      \begin{center}  
        \subfigure[Reply redundancy]{
          \label{fig:dup-scan-angle}
          \includegraphics[scale=0.3]{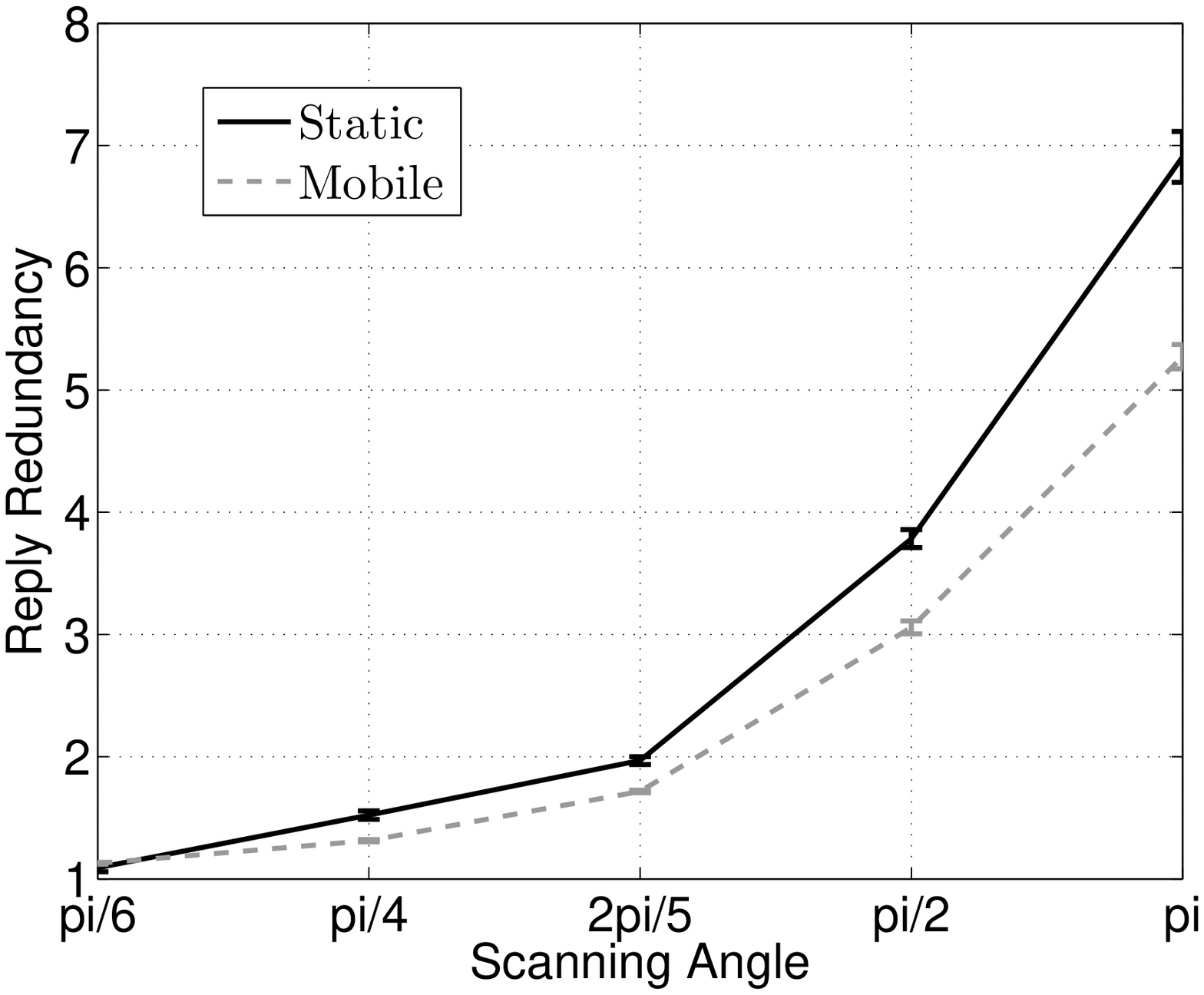}}
      \end{center}
    \end{minipage}

   &

    \begin{minipage}[t]{0.3\textwidth}
      \begin{center}    
        \subfigure[Latency]{
          \label{fig:delay-scan-angle}
          \includegraphics[scale=0.3]{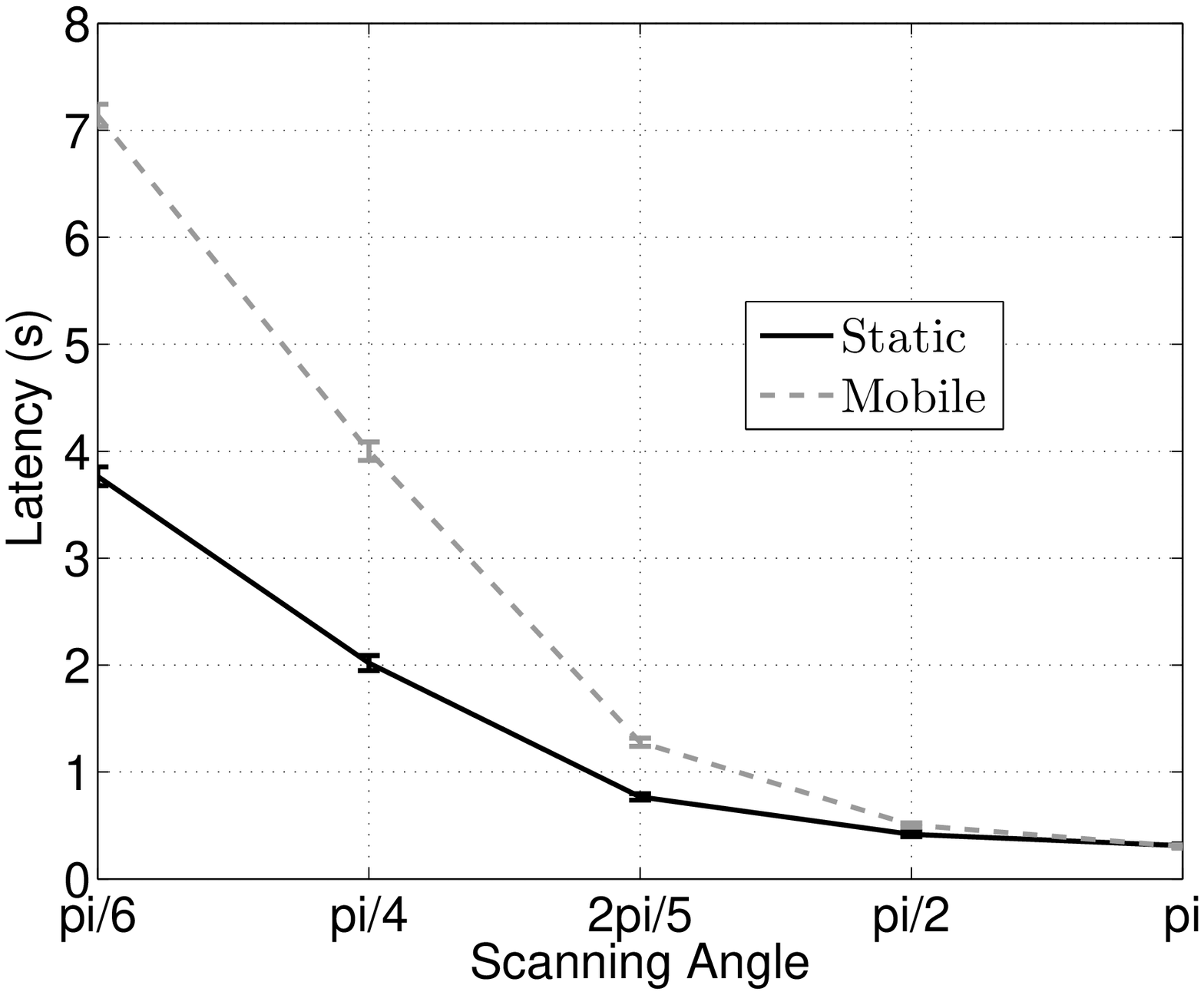}}
      \end{center}
    \end{minipage}
    
  \end{tabular}
  \caption{Performance of the scanning mechanism as a function of the scanning angle (sector timeout = 0.5s, $|\mathcal{C}|$=30 and $\tau$=100 s).}
  \label{fig:performance-scan-angle}
\vspace{-3mm}
\end{figure*}

Fig.~\ref{fig:performance-scan-angle} shows the impact of the number of angular
sectors in which the space around a node is partitioned, as determined by the
scanning angle parameter. As explained in 
Sec.~\ref{sec:content-access}, a small scanning angle might reduce the
probability for a query to reach a replica, hence the lower solving ratio
with small angles. We observe a similar effect on redundancy: smaller
angles limit the number of replicas ``hit'' by a query. Instead, the latency
decreases with larger scanning angles because the probability to find a 
replica within a sector increases.

%As a general comment on the impact of node mobility, both
%Fig.~\ref{fig:performance-scan-time} and \ref{fig:performance-scan-angle}
%indicates that in a dynamic network the delay for a node to access and retrieve
%information deteriorates.

\vspace{10pt}

\textit{ \textbf{Summary:} we analyzed the performance of several content access
mechanisms, ranging from simple flooding-based to complex schemes requiring
perfect-discovery. With the setting used in our tests,
we showed that a content query hits at least one replica with very high 
probability (Fig. \ref{fig:performance_bis})
and that access delay can be slightly larger than 1 s with the scanning
mechanism. Despite having larger delays, our results 
(Figs.~\ref{fig:performance-scan-time}
 and~\ref{fig:performance-scan-angle}) showed that the scanning
mechanism achieves very low redundancy (comparable to perfect-discovery) and
bears little costs in terms of complexity (which is comparable to flooding). }

\subsubsection*{Is the replication mechanism effective in reaching a target number of replicas?}

We now turn our attention to the  \textit{uncapacitated facility
location} described in Sec.~\ref{sec:background} and study how well the
replication mechanism defined in Sec.~\ref{sec:content-replication} approximates
the joint problem of replication and placement. Recall that the 
hypothesis of content demand points to be associated to the closest facility is
hardly viable in the context of broadcast wireless networks
(only the perfect-discovery scheme achieves it but with significant additional
complexity).

Here we take an extreme scenario in which only one copy of the content is
initially present in the network and we focus on the evolution in time of the
number of replicas in the system. We omit the temporal evolution of the $\chi^2$ index, 
since our results are consistent with what we have observed
for the placement scheme without replication. 

\begin{figure}[htbp]
\begin{center}
  \includegraphics[scale=0.35]{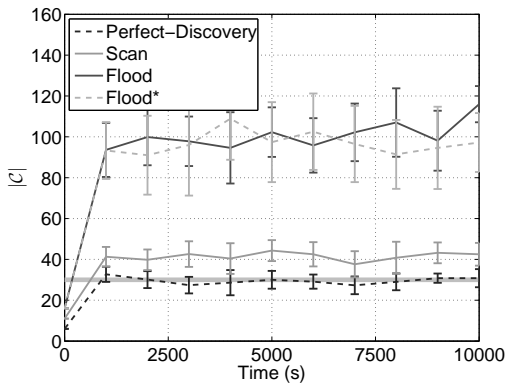}
\end{center}
\caption{Temporal evolution of the number of replicas, 
for a network bootstrapping with $|\mathcal{C}|=1$ in a mobile scenario 
($\lambda=0.01$, $s_{R}=10$, $\tau=100$ s, $|\mathcal{C}^*|=30$).\label{fig:temporal_evo}}
\vspace{-3mm}
\end{figure}

Fig.~\ref{fig:temporal_evo} shows the temporal evolution of the total number of
content replicas $|\mathcal{C}|$ for the mobile scenario. In the plot, we
report a reference line representing the target number of content replicas
computed as follows. Let us consider the perfect-discovery content access
mechanism: demand points are associated to their closest replica. Finding the
optimal number of content replicas amounts to solving the uncapacitated facility
location problem for a given network graph. We have implemented the
\textit{centralized} algorithm in \cite{arya01}\footnote{We set a non-uniform cost to
open a facility proportional to its degree: indeed, a highly connected node will
most likely attract more demand from content consumers.}, which is 
a rigorous but very demanding approach when the network is dynamic, despite a polynomial execution
time. The optimal number of content replicas, however, can be approximated:
assume each replica node to receive, on average, the same share of queries.
Then, we would like the target number of replica nodes $|\mathcal{C}^*|$ 
to be such that:

\begin{equation}
\label{eq:total-load}
|\mathcal{C}^*| \frac{s_{R}}{\tau} = (N-|\mathcal{C}^*|)\lambda
\end{equation}
where $(N-|\mathcal{C}^*|)\lambda$ is the network aggregate query rate. 
From (\ref{eq:total-load}), we write:
\begin{equation}
\label{eq:ideal-c}
|\mathcal{C}^*| = \frac{N\lambda\tau}{\lambda\tau + s_{R}} 
\end{equation}
For the parameters used in our simulations, 
the sample solution of the centralized algorithm and the value in 
(\ref{eq:ideal-c}) 
agree on the target number of replicas: the system 
should reach $|\mathcal{C}^*|=30$ replica nodes. 

%{\color{red}In mobile scenarios we use our approximation as a reference value: we sampled the reference over several time-instants to verify its consistency compared to the optimal solution.}

Fig.~\ref{fig:temporal_evo} indicates that the number of content replicas we
achieve with our scheme strikingly matches the target value when
perfect-discovery is used: in steady state, the average relative error is less
than 2\%. We can also observe the adverse effects of the artificially inflated
workload created by the scanning and flooding mechanisms. In case of flooding,
the number of replicas in the system is drastically overestimated. Despite its
simplicity, the scanning mechanism induces a much smaller error than flooding in
reaching the target number of content replicas.

\begin{figure}[htbp]
\begin{center}
  \includegraphics[scale=0.3]{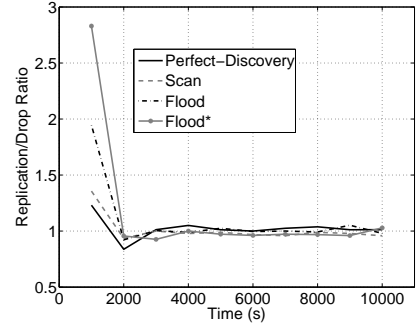}
\end{center}
\caption{Replication/drop ratio for a mobile scenario for 
a network bootstrapping with $|\mathcal{C}|=1$  
($\lambda=0.01$, $s_{R}=10$, $\tau=100$ s).\label{fig:repdrop_ratio}}
\end{figure}

Fig.~\ref{fig:repdrop_ratio} depicts the ratio between the aggregate
number of replication and drop decisions\footnote{For sake of clarity, in this
figure we omit handover decisions and we report results every 1000 s.} and
shows in more details the behavior of the replication mechanism: when a
single or few nodes support most of the content queries, the number of
replication decisions is considerably higher than drop decisions. Once a
sufficient number of replicas have populated the network, the ratio
reaches the steady state value of 1. 

\subsubsection*{How is the total workload shared among replica nodes?}

As before, we study the joint placement and replication problem and we use the
extreme scenario in which the network is initialized with only one content
replica. 

\begin{figure}[htbp]
\begin{center}
  \includegraphics[scale=0.3]{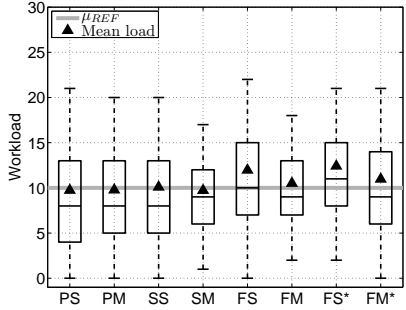}
\end{center}
\caption{Aggregate workload distribution for replicas for a 
network bootstrapping with $|\mathcal{C}|=1$ ($\lambda=0.01$, $s_{R}=10$, $\tau=100$ s).\label{fig:loadbp}}
\vspace{-3mm}
\end{figure}

Fig.~\ref{fig:loadbp} shows the 25\%, 50\% and 75\% quantiles of the workload
for each replica node, aggregated over the simulation time. The figure is
complemented with the \textit{average} workload per replica node. The reference
value $s_{R}=10$ is shown as a horizontal line in the plot. As expected, with
perfect-discovery the average load matches the reference value $s_{R}$, both in
the static and mobile scenario. Instead, when flooding is used, the average load
is consistently above $s_{R}$, albeit the ``approximation error'' is smaller
than 3\%. It should be noted, with reference to Fig.~\ref{fig:temporal_evo},
that, when flooding is used, replica nodes achieve a good approximation of $s_{R}$ because the total
number of content replicas is higher than the target value: since the workload
induced by flooding reaches multiple replica nodes, our adaptive scheme
replicates more than necessary to maintain the workload within the desired
target $s_{R}$. When scanning is used, the quality of approximation of $s_{R}$
is excellent. 
%better than what achieved by flooding. 
Node mobility helps in reducing the skewness of the workload distribution. We note that it is possible that some
nodes experience very little workload: this happens when the location of a
replica is very close to the boundaries of the node deployment area, which
implies that fewer content queries will reach those replica nodes. 

\subsubsection*{What is the convergence time of the replication mechanism?}

Convergence time should be carefully defined in our context: clearly, our
mechanism cannot settle to a static, unique content replica placement, nor can
it stabilize on a unique number thereof. For placement, it is not our intent to
statically assign the role of content replica to a node and deplete nodal
resources: we seek to balance the workload across all network nodes. We assume
the network to have converged to a steady state when the difference between the
reference value computed through (\ref{eq:ideal-c}) 
and the experimental number of replicas is within 2\%.

Again, we consider a \textit{worst-case} scenario in which only one copy of the
content is initially present in the network. Tab.~\ref{tab:ts} illustrates how
convergence time (labelled $t_s$) varies with the storage time $\tau$, and the
tolerance value $\epsilon$ in the  case of perfect-discovery. We also performed
experiments to study the impact of the network size: we have observed a linear
growth of the convergence time with $N$. Since the storage time $\tau$ is used
to trigger replication/drop decisions, we expect to see a positive
correlation between $\tau$ and convergence time: Tab.~\ref{tab:ts} confirms this
intuition. We note that there is a trade-off between the convergence time, and
the message overhead: a small storage time shortens the convergence time at the
cost of an increased number of content movements from a node to another.
Moreover, in our simulations we do not trace the message overhead required by
the perfect discovery mechanism: with frequent reassignments of a node to the
closest replica, this overhead could become prohibitive. As for the impact of
the tolerance parameter $\epsilon$, our experiments
indicate that a very reactive scheme would yield smaller
convergence times, at the risk of causing frequent oscillations around a 
target value.

\begin{table}[htbp]
  \centering
 \caption{Average convergence time $t_S$ as a function of the storage time $\tau$ 
($\epsilon=2$) and the tolerance factor $\epsilon$ ($\tau=100$ s), with perfect-discovery.\label{tab:ts}}
  
  \begin{tabular}{cc}

    {
      \begin{tabular}{|c|c|}
        \hline
        $\tau$ (s) & $t_S$ (s) \\
        \hline
        \hline
        {20} & {800} \\
        \hline
        {50} & {1500} \\
        \hline
        {100} & {1700} \\
        \hline
        {150} & {2000} \\
        \hline
        {200} & {2300} \\
        \hline
      \end{tabular}      
    } 
    
    & 

    {
      \begin{tabular}{|c|c|}
        \hline
        $\epsilon$ & $t_S$ (s) \\
        \hline
        \hline
        {0} & {700} \\
        \hline
        {2} & {1700} \\
        \hline
        {5} & {1900} \\
        \hline
      \end{tabular}
    }
  \end{tabular}
\vspace{-3mm}
\end{table}

\vspace{10pt}

\textit{ \textbf{Summary:} we showed that our replication mechanism, 
which provides for
content copies to by added, dropped or handed over by replica nodes achieves a very
good approximation of the optimal number of replicas (Fig. \ref{fig:temporal_evo}) and a target placement
thereof (Fig. \ref{fig:loadbp}), in a variety of scenarios and under several content access mechanisms
(Fig. \ref{fig:repdrop_ratio}) .
Our scheme is robust against mobility, which turns out to be an ally especially
for balancing the workload among replica nodes. We also studied the parameters
that influence the time it takes for the network to reach a steady state and
discussed the tradeoff that exists between convergence time and message
overhead (Tab. \ref{tab:ts}). The scanning mechanism suggested in this work mitigates the message
overhead due to an auxiliary service to support perfect discovery, at the
cost of increasing the estimation error in the number of content replicas to
place in the network.  }

\subsubsection*{What is the impact of variations in time and in space of the content demand?}

We now focus our attention on the behavior of content replication in presence of
a dynamic workload. We first examine workload variations in time. In a first
phase, that begins at time 0 and ends at 5000 s, we set the content request
rate as $\lambda=0.01$ req/s. In a second phase, from 5000 s to the end of the
simulation, the request rate doubles, i.e., $\lambda=0.02$ req/s.

\begin{figure}[htbp]
  \begin{center}    
    \includegraphics[scale=0.3]{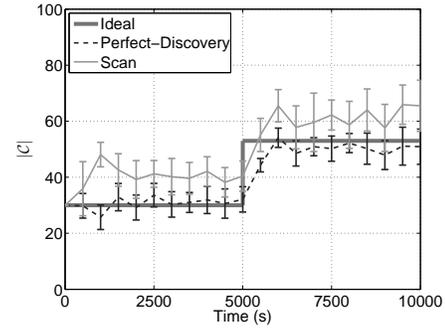}
  \end{center}
  \caption{\label{fig:time_variation} Temporal evolution of the number of replica nodes in case 
of variations in time of the content demand, for a mobile network. $|\mathcal{C}^*|$ is equal 
to 30 and 53 in the first and second phase, respectively.}  
\vspace{-3mm}
\end{figure}

\begin{figure}[htbp]
      \begin{center}    
         \includegraphics[scale=0.3]{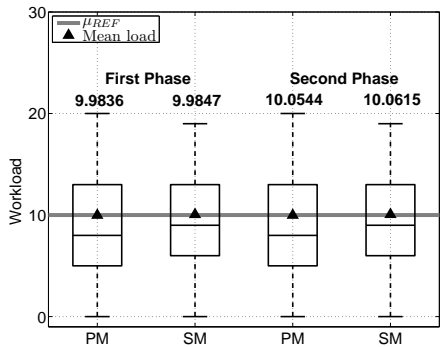}
      \end{center}
  \caption{Workload distribution of replica nodes for variations in time of the content demand, 
in a mobile network.\label{fig:load_variation}}
\vspace{-3mm}
\end{figure}

Fig.~\ref{fig:time_variation} shows the temporal evolution of the number of
replicas in a mobile network. The figure is enriched with two reference
values\footnote{As explained before, we compute the target values through
(\ref{eq:ideal-c}), but compare them with the solution of the facility location problem computed over
several snapshots of the network graph.}: 
in the first phase $|\mathcal{C}^*|=30$,
in the second phase $|\mathcal{C}^*|=53$. We report results for the
perfect-discovery and the scanning access schemes: our mechanism achieves a very
good approximation of the target number of replicas with perfect-discover, and
slightly over-replicates the content when scanning is used. Despite node
mobility, not only is our scheme able to correctly determine the number of
replicas but also their target location; as a consequence, the load distribution
is minimally affected by a variation in time of content demand.
This result is shown in Fig.~\ref{fig:load_variation}, where we indicate the
25\%, 50\% and 75\% quantiles of the workload, and we report the average load per
replica node. Note that, although at first glance the mean load values look 
identical, there is a minimal difference among them, as shown by the numbers
above the error bars in the plot. 

We now turn our attention to variations in space of content demand: we describe
the behavior of the content replication mechanism with the following example.
For the initial $5000$~s of the simulation time, content queries are
issued by all nodes deployed on the network area $\mathcal{A}$ of size $200$
m$^2$. Subsequently, we select a smaller square area $\alpha$ of size $100$
m$^2$ and instruct only nodes within that zone to issue content queries, while
all other nodes exhibit a lack of interest. Nodes use perfect-discovery to access
the content.

\begin{figure}[htbp]
  \begin{center}
     \includegraphics[scale=0.3]{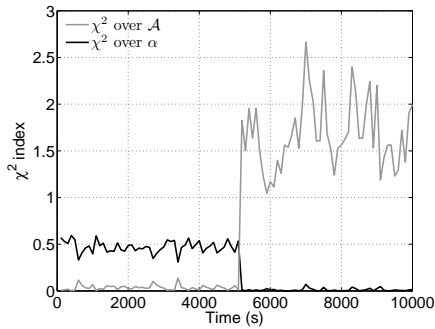}
  \end{center}
  \caption{Temporal evolution of the $\chi^2$ index for variation in space 
of the content demand, in a mobile network. \label{fig:load_in_space}}
\vspace{-3mm}
\end{figure}

Fig.~\ref{fig:load_in_space} shows the temporal evolution of the $\chi^2$ index
computed as follows. We compare the distribution of content replicas achieved by
our mechanism against the target \textit{nodal uniformity} distribution
computed over the network area $\mathcal{A}$ and the sub-zone $\alpha$, and plot
the two curves. When all nodes issue queries, the line of the $\chi^2$ index
computed over $\mathcal{A}$ indicates a good approximation of the target replica
placement. This is true up to roughly 5000 s, after which the $\chi^2$ in
$\mathcal{A}$ shows a substantial deterioration in approximating nodal
uniformity on all nodes. Indeed, after 5000 s, the target replica placement
should be computed over the area $\alpha$, and the second line of the $\chi^2$
index computed over $\alpha$ indicates that the distribution of replicas
achieved by our mechanism is indeed a good approximation of a target placement
over $\alpha$. We can conclude that our mechanism allows content replicas to
``migrate'' where the demand is higher.

\vspace{10pt}

\textit{ \textbf{Summary}: we showed that our mechanism achieves a target
placement and a sufficient number of content replicas to cope with complex
demand scenarios, even in a mobile network. When content demand varies in time,
our mechanism adaptively replicates the content to meet the variation in the
workload (Fig. \ref{fig:time_variation}). When content demand varies in space, our scheme allows content
replicas to migrate to the location where the demand is higher and 
meet a variation in the workload (Fig. \ref{fig:load_in_space}).  }

\section{Related work on replication in multihop networks}
\label{sec:related-work}

%Several works have dealt with replication
%in mobile networks; here, we review the ones that are more  
%relevant to our study, and we highlight the differences with 
%respect to our solution.

Simple, widely used techniques for replication are gossiping
and epidemic dissemination \cite{HaraEpidemic,Simplot}, where the information is forwarded to a randomly
selected subset of neighbors. Although our RWD scheme may resemble
this approach in that a replica node hands over the content to a
randomly chosen neighbor, the
mechanism we propose and the goals it achieves (i.e., approximation of
nodal uniformity and optimal number of replicas)
are significantly different.

Another viable approach to replication is represented
by quorum-based \cite{Hubaux} and cluster-based protocols \cite{Hassanein05}.
Both methods, although different, are based
on the maintenance of quorum systems or clusters,
which in mobile network are likely to cause an exceedingly
high overhead.
Node grouping is also exploited in \cite{Hara01,Hara03},
where groups of nodes with stable links are used
to cooperatively store contents and share information.
The schemes in \cite{Hara01,Hara03}, however, require
an a-priori knowledge of the query rate, which is 
assumed to be constant in time.
Note that, on the contrary,  our lightweight solution
can cope with a dynamic demand, whose estimate by the
replica nodes is used to trigger replication.
We  point out that achieving content diversity is  
the goal of \cite{Yin} too, where, however, cooperation is exploited 
among one-hop neighboring nodes only.

Threshold-based mechanisms for content replication
are proposed in \cite{Almeroth,Hara0457}. In particular,
in \cite{Almeroth} it is the original server that 
decides whether to replicate content or not, and 
where. In \cite{Hara0457},  nodes have limited storage
capabilities: if a node does not have enough free memory,
it will replace a previously received content with a new one, only
if it is going to access that piece of information more frequently
than its neighbors up to $h$-hops. Our scheme 
significantly differs from these works, since it is a totally distributed
and extremely lightweight mechanism, which accounts for 
the content demand by other nodes and ensures 
a replica density that autonomously adapts to the 
changes in the query rate over time and space.

Finally, relevant to our study are the numerous schemes proposed
for handling query/reply messages; examples are \cite{Chen},
which resembles the perfect-discovery mechanism, and 
 \cite{Estrin02,Vaidya} where queries are propagated
along trajectories so as to meet the requested information. 
Also, we point out that the RWD scheme
was first proposed in our work \cite{casetti09}.
That paper, however, besides being a preliminary
study, focused on
mechanisms for content handover only: no
replication or content access were addressed.

\section{Conclusions}

We focused on content replication in mobile networks and we
addressed the joint optimization problem of (i) establishing the number of
content replicas to deploy in the network, (ii) finding their most suitable
location, and (iii) letting users efficiently access content through
device-to-device communications.

To achieve these goals, we proposed a distributed mechanism that lets content
replicas move in the network according to random patterns: network nodes
temporarily store content, which is handed over to randomly selected neighbors.
Hence the burden of storing and providing content is evenly shared among nodes
and load balancing is achieved. In our mechanism, replica nodes are also
responsible for creating content copies or drop them, with the goal of obtaining
an ideal number of content replicas in the network. The workload experienced by
a replica node is the only measured signal we use to trigger replication and
drop decisions. 

We studied the above problems through the lenses of facility location theory and
showed that our lightweight scheme can approximate with high accuracy the
solution obtained through centralized algorithms. Clearly, network dynamics
exact a high toll in terms of complexity to reach an optimal replication and
placement of content, and we showed that our distributed mechanism can readily
cope with such a scenario. Moreover, we removed the typical assumption of
assigning content demand points to their closest replica and investigated
several content access schemes, their performance, and their impact on content
replication. 

Lastly, we studied the flexibility of our scheme when content demand varies
in time and in space: our experiments underlined the ability of our approach to
adapt to such variations while maintaining accuracy in approximating an optimal
solution.

Our next step will be to relax the assumption of a cooperative setting and
analyze selfish replication with tools akin to game theory. In \cite{podc09} we
show that the system we study can be modeled as an anti-coordination game, and
our goal is to understand how to modify or extend the ideas presented in this
work to achieve strategy-proofness.

\begin{small}

\end{small}

\end{document}